\documentclass[10pt,twocolumn,twoside]{IEEEtran}
\IEEEoverridecommandlockouts                            

\usepackage{cite}
\usepackage{url}
\usepackage{amsmath,amssymb,amsfonts}
\usepackage{algorithmic}
\usepackage{graphicx}
\usepackage{textcomp}
\usepackage{balance}
\usepackage{xcolor}
\usepackage[utf8]{inputenc}

\usepackage{flushend}

\newcommand{\iRipple}{I_{\mathrm{ripple}}}
\newcommand{\iRippleW}{I_{\mathrm{ripple},\,\omega}}
\newcommand{\iRippleWw}{I_{\mathrm{ripple},\,2\omega}}
\newcommand{\iRippleMax}{\iRipple^{\mathrm{Max}}}
\newcommand{\pRippleWw}{P_\mathrm{ripple,\,2\omega}}
\newcommand{\Iloading}{I_\mathrm{loading}}
\newcommand{\IloadingPhi}{I_\mathrm{loading,\,\phi}}
\newcommand{\IloadingPhiT}{I_\mathrm{loading,\,\phi,\,\tau}}
\newcommand{\Idemand}{I_\mathrm{demand}}
\newcommand{\Vdclink}{v_\mathrm{dc\,link}}
\newcommand{\IhalfBridge}{I_{\mathrm{HB}}}
\newcommand{\IhalfBridgeMax}{I_{\mathrm{HB}}^{\mathrm{Max}}}
\newcommand{\IhalfBridgePhi}{I_{\mathrm{HB},\,\phi}}
\newcommand{\IhalfBridgePhiMax}{I_{\mathrm{HB},\,\phi}^{\mathrm{Max}}}
\newcommand{\IhalfBridgeTotal}{I_{\mathrm{HB}}^{\mathrm{Total}}}
\newcommand{\Vgrid}{V_{\mathrm{grid}}}
\newcommand{\VgridPhi}{V_{\mathrm{grid,\,\phi}}}
\newcommand{\VgridPhase}{\Vgrid^{\mathrm{phase}}}
\newcommand{\Resr}{r}

\graphicspath{{./figures/}}
\makeatletter
\def\input@path{{./tables/}}
\makeatother

\def\BibTeX{{\rm B\kern-.05em{\sc i\kern-.025em b}\kern-.08em
    T\kern-.1667em\lower.7ex\hbox{E}\kern-.125emX}}
\usepackage{lipsum}

\title{\LARGE \bf 
DC Link Capacitor Ripple Constraints Limit the Benefits\\of Utility-Owned Four-Wire Power Converters
}

\author{Matthew Deakin, Xu Deng, Shafiq Odhano and Rahmat Heidari
\vspace{-0.5cm}
 \thanks{
M. Deakin, S. Odhano, X. Deng are with the School of Engineering, Newcastle University, Newcastle-upon-Tyne, United Kingdom. R. Heidari is with the School of Electrical Engineering and Computer Science, University of Queensland, Brisbane, Australia. M. Deakin was supported by the Royal Academy of Engineering under the Research Fellowship programme. Email: {\tt matthew.deakin@newcastle.ac.uk}}
}

\begin{document}
\begingroup
\allowdisplaybreaks

\maketitle

\begin{abstract}
Utilities are increasingly interested in power converters to increase the headroom of their assets by actively controlling power flows on their network. In this work we demonstrate that thermal limits of dc link capacitors can result in substantially diminished benefits of these converters under unbalanced operation, due to constraints on neutral current and double-line frequency power ripple. Considering nine voltage source converter topologies with varying ripple capabilities, the upper bound (in terms of additional headroom released) increases by more than 80\% compared to a no-ripple case for the application of phase current unbalance mitigation.
\end{abstract}


\section{Introduction}

Power converters are achieving ever-increasing power density and cost-effectiveness, driven by a huge range of applications including electric vehicles, motor drives, and battery storage systems. Distribution system operators are therefore exploring how these next-generation power converters could be exploited to avoid disruptive and costly traditional reinforcement as electrification continues to gather pace. Power converter-based technologies which are being explored by utilities include back-to-back soft open points, STATCOMs, and hybrid transformers.

As a result of these new applications, there has recently been interest in more accurate algebraic modeling of power converters. Aspects considered include physics-based loss modeling \cite{badmus2026twostage}, consideration of controls on power quality impacts of power converters \cite{heidari2024improved}, or dc link capacitor ripple on operational constraints under unbalanced operation \cite{ziyat2023voltage, deakin2025power}. 

This work focuses on the latter consideration of low-frequency dc link ripple (e.g., ripple components with frequency $<50\omega$ rad/s, for fundamental grid angular frequency $\omega$) under the injection of unbalanced currents. Low-frequency dc link ripple is caused by neutral return current, and double line frequency power oscillation as a result of time-varying active power injection.

The contribution of the work is to use analytic bounds on minimum and maximum dc link ripple values to conduct a parameter sweep that evidence, for a range of common voltage source converter (VSC) topologies, that dc link ripple substantially and consistently impacts on the benefits of VSCs for applications requiring unbalanced injections. As compared to recent works \cite{ziyat2023voltage, deakin2025power}, we consider benefits for a variety of topologies, considering the full range of capacitor sizes.

\paragraph{Notation} Upper case variables represent phasors. Superscript $(\cdot )^{*}$ represents complex conjugate and $|\cdot|$ the magnitude of a complex number. The (angular) line frequency of the grid is $\omega$. $\phi$ is used as a (subscripted) index for vectors which collect per-phase (or per-leg) quantities.

\section{Voltage Source Converter Modelling with\\DC Link Ripple Constraints}

Fig.~\ref{f:model} shows a model of a grid-connected four-wire, four-leg VSC with split dc link (the parallel fourth leg and split dc link splits the neutral current to reduce stress on each of those return paths). There are a range of topologies for the filter block (see, e.g., \cite{blas2023sic}): for the purposes of this work, we assume the filter is primarily introduced to mitigate high-frequency ripple components, with little voltage drop or current transfer between phases, so the voltage drop between the half-bridges and grid at low frequency is negligible. Similar topologies to Fig.~\ref{f:model} are also common, such as three-leg, four-wire topologies (i.e., with only three half bridges \cite{ziyat2023voltage}), or for four-leg four-wire topologies with monolithic dc link (i.e., the dc link has only a single capacitor between upper and lower rails, with no midpoint \cite{deakin2025power}). 

\begin{figure}
\centering
\includegraphics[width=0.46\textwidth]{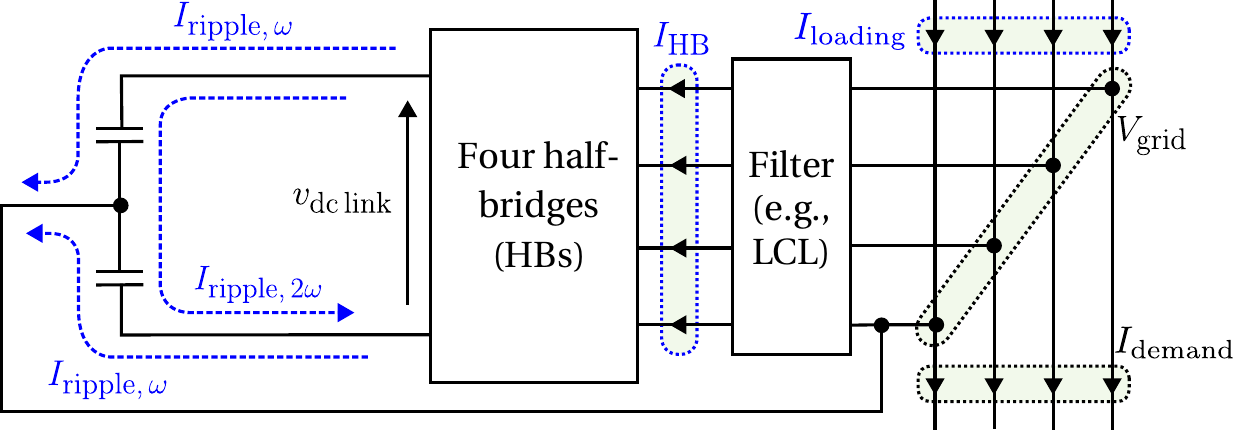}
\caption{A four-leg, four-wire VSC converter with split dc link. Quantities in \textcolor{blue}{blue} are variables considered to change significantly with half bridge currents $\IhalfBridge$ assigned by the VSC controller. For the purposes of this paper, the impact of the filter on current and voltage at line frequency is neglected, so the voltage and current at the input and output of the filter are equal.}\label{f:model}
\end{figure}

\subsection{Basic Voltage Source Converter Constraints}
For any four-wire VSC topologies, current injections into the half bridge legs, $\IhalfBridge$, must be limited according to the thermal rating of semiconductor devices $\IhalfBridgePhiMax$, i.e.,
\begin{equation}\label{e:i_leg_max}
|\IhalfBridgePhi| \leq \IhalfBridgePhiMax \quad \forall \,\, \phi \in \Phi \,,
\end{equation}
where $\Phi$ represents the set of half-bridge legs. If there is no dc-side source capable of providing active power, then
\begin{equation}\label{e:p_balance}
\sum _\phi \VgridPhi \IhalfBridgePhi ^* = 0\,,
\end{equation}
where $\Vgrid$ is the vector of grid voltages at the point of common coupling. For the purposes of this work, this grid voltage is assumed to be dominated by a nominal positive sequence voltage only, i.e.,
\begin{equation}\label{e:grid_voltage}
\Vgrid = \Vgrid^{\mathrm{phase}}\,[1,\,\alpha^{-1},\, \alpha^{-2},\,0\,]^{\intercal}\,,
\end{equation}
where $\alpha=e^{2\pi/3}$ is the 120$^{\circ}$ phase rotation operator and $\Vgrid^{\mathrm{phase}}$ is the nominal phase voltage.

\subsection{Voltage Source Converter Capacitor Ripple Constraints}
For VSCs injecting line-frequency currents, low-frequency ($<50\omega$) dc link (capacitor) ripple is dominated by line frequency neutral return current and double-line frequency power ripple. Line frequency neutral current occurs only when the dc link has a split-capacitor topology, so that neutral current can return through the midpoint between the capacitors (shown as $\iRippleW$ on Fig.~\ref{f:model}). Assuming dc link capacitors of equal capacitance and balanced charge, then the line-frequency neutral current ripple is shared between capacitors equally \cite{ziyat2023voltage},
\begin{equation}\label{e:i_ripple_w}
\iRippleW = \dfrac{1}{2} \sum _\phi \IhalfBridgePhi \,.
\end{equation}

Double-line frequency power ripple occurs as a result of non-cancellation of active power ripple across phases under unbalanced power injections. The general form for this power ripple $\pRippleWw$ is \cite{deakin2025power}
\begin{equation}
\pRippleWw = \left  |\sum _\phi \VgridPhi \IhalfBridgePhi \right | \,,
\end{equation}
so, for a known dc link voltage $\Vdclink$, the current ripple is
\begin{equation}\label{e:i_ripple_2w}
|\iRippleWw | = \dfrac{ \left |\sum _\phi \VgridPhi \IhalfBridgePhi \right |}{\Vdclink}\,.
\end{equation}

The impact of ripple currents on the thermal load of the capacitor is modeled via its equivalent series resistance (ESR), $\Resr$, so that for each harmonic $k$,
\begin{equation}
\begin{split}
\label{e:esr_defn}
\mathrm{Thermal\,Load} &= \Resr \sum _k |I_{\mathrm{ripple}}(k\omega)|^2 \\
	&= c_{\mathrm{Sw.}} + \Resr \left (
		|\iRippleW |^2 + |\iRippleWw |^2
\right ) \,,
\end{split}
\end{equation}
where $c_\mathrm{Sw.}$ represents losses due to high-frequency (low-energy) switching ripple from high frequency half-bridge pulse width modulation (PWM), and the line and double-line frequency components $\iRippleW,\,\iRippleWw$ are from \eqref{e:i_ripple_w}, \eqref{e:i_ripple_2w}. For a given maximum thermal load, the maximum allowed ripple currents through the capacitor can be limited. Defining
\begin{equation}
(\iRippleMax ) ^2 = \dfrac{\mathrm{Thermal\,Limit} - c_{\mathrm{Sw.}}}{r}
\end{equation}
then assigning Thermal Limit as the upper bound for \eqref{e:esr_defn} yields
\begin{equation}\label{e:ripple_thermal_constraint}
|\iRippleW |^2 + |\iRippleWw |^2  \leq (\iRippleMax ) ^2\,.
\end{equation}
The voltage ripple at a given frequency $\omega$ can be determined according to the fundamental capacitor relation $C=Q/V$ for the known current ripple, with constraints included for each frequency \cite{ziyat2023voltage}. For the remainder of this work, we focus on thermal constraints of the form \eqref{e:ripple_thermal_constraint}.

\subsection{Maximum DC Link Ripple Current Design}

Considering \eqref{e:ripple_thermal_constraint}, it is possible to determine an upper bound on the maximum theoretical ripple capacity $\iRippleMax$ for a given converter leg capacity $\IhalfBridgeMax$. Considering the grid voltage \eqref{e:grid_voltage}, the negative sequence current injection is the only injection which will increase the value of the current ripple \eqref{e:i_ripple_2w}. The maximum value of the double-line frequency current ripple $\iRippleWw$ will take value of the kVA rating of the converter divided by the dc link voltage. For a sinusoidal pulse width modulation (SPWM) scheme, the dc link voltage should be a minimum of $2\sqrt{2}$ that of $\VgridPhase$; a dc link voltage $\Vdclink=3\VgridPhase$ has headroom of 6\%. Therefore, noting that a fourth leg does not contribute to active power ripple (as it is at 0~V, from \eqref{e:grid_voltage}), and considering \eqref{e:i_leg_max} and \eqref{e:i_ripple_2w}, the maximum value of $|\iRippleWw|$ takes value
\begin{equation}\label{e:monolithic_cap_ripple}
|\iRippleWw | \leq \dfrac{1}{3} \sum _{\phi=\{1,\,2,\,3\}} \IhalfBridgePhiMax \,,
\end{equation}
for any permissible value of $\IhalfBridge$ \eqref{e:i_leg_max}. 

The maximum line-frequency ripple $|\iRippleW|$ occurs when all legs inject currents $\IhalfBridge$ with identical phase angle, so for any $\IhalfBridge$,
\begin{equation}\label{e:neutral_ripple}
|\iRippleW | \leq \dfrac{1}{2} \sum _\phi \IhalfBridgePhiMax \,.
\end{equation}
To consider the maximum ripple $\IhalfBridgeMax$ across all $\IhalfBridge$, note that \eqref{e:ripple_thermal_constraint} can considered a weighted squared 2-norm of the zero- and negative-sequence current injections (from \eqref{e:monolithic_cap_ripple}, \eqref{e:neutral_ripple}). However, given that sequence components are orthogonal, the maximum ripple current $\iRippleMax$, is the maximum of the individual $\omega,\,2\omega $ ripple currents. As a result, we can write down that
\begin{equation}\label{e:ripple_sweep_parameters}
0 \leq \iRippleMax \leq \dfrac{1}{2} \sum _\phi \IhalfBridgePhiMax\,.
\end{equation}
Given \eqref{e:ripple_sweep_parameters}, it is possible to consider how dc link capacitors impact on converter benefits across the full range of reasonable capacitor thermal capacities.

\subsection{Operational Constraints for three-leg, four-leg, and reconfigurable VSCs}\label{ss:vsc_operational_constraints}
The operating constraints considered in this work are for three different half-bridge configurations, as follows.

\paragraph{Three-leg VSC} The VSC's operational constraints are given by \eqref{e:i_leg_max}, \eqref{e:p_balance}, \eqref{e:grid_voltage}, \eqref{e:i_ripple_w}, \eqref{e:i_ripple_2w}, and \eqref{e:ripple_thermal_constraint}, with the set of leg phases $\Phi = \{1,\,2,\,3\}$. All three legs have the same capacity, $\IhalfBridgePhiMax = \IhalfBridgeTotal/3$.

\paragraph{Four-leg VSC} A four-leg VSC has the same operational constraints as for a three-leg VSC, but has $\Phi = \{1,\,2,\,3,\,4\}$. Each leg is assumed to have the same current carrying-capacity, $\IhalfBridgePhiMax = \IhalfBridgeTotal/4$, with the fourth leg enabling reduced line frequency (neutral) current ripple $\iRippleW$.

\paragraph{Reconfigurable VSC} Finally, to consider other possible half-bridge capacity sizes, we consider the VSC leg capacity convex relaxation introduced for reconfigurable VSCs \cite{deakin2023multiplexing}. This relaxation provides a continuous approximation of a (discrete) power multiplexer that enables the output of half bridges to be connected to different phases \cite{deakin2024reconfigurable}; therefore, on a per-unit basis, this relaxation provides an upper bound to VSC performance under any leg sizing. This topology still considers constraints \eqref{e:p_balance}, \eqref{e:grid_voltage}, \eqref{e:i_ripple_w}, \eqref{e:i_ripple_2w}, and \eqref{e:ripple_thermal_constraint}, but replaces \eqref{e:i_leg_max} with the convex relaxation
\begin{equation}
\sum \IhalfBridge \leq \IhalfBridgeTotal \,.
\end{equation}

\subsection{Application: Phase Current Balancing}

Typically, thermal constraints on distribution system assets are binding on a per-phase basis. Therefore, by balancing the unbalanced demand $\Idemand$ (imbalance caused by, for example, single-phase domestic loads), the per-phase asset loading $\Iloading$ can be reduced, increasing the headroom of a distribution system asset. The benefit of a VSC at a time $\tau$ for a VSC is
\begin{equation}
\label{e:benefit}
\mathrm{Benefit}_\tau = \left ( \max_\phi |\IloadingPhi^{\mathrm{No\,VSC}}| \right ) - \left (\max_\phi |\IloadingPhi^{\mathrm{With\,VSC}}| \right )\,.
\end{equation}
For demand $\Idemand$ at time $\tau$, the objective is to minimize the maximum phase current (i.e., maximising the benefit \eqref{e:benefit}), as
\begin{equation}\label{e:objective}
\min _{\IhalfBridge} \quad \left (\max_\phi |\IloadingPhi^{\mathrm{With\,VSC}}| \right ) + \lambda \| \IhalfBridge \|_2\,,
\end{equation}
with constraints on $\IhalfBridge$ described in Section~\ref{ss:vsc_operational_constraints} for each of the three VSC topologies considered; regularization parameter $\lambda \ll 1$ ensures a unique solution in $\IhalfBridge$ is found (and can be considered a proxy as a secondary objective to minimize converter losses).

By solving \eqref{e:objective} at each time period $\tau$, the additional headroom can be found using each solution $\IhalfBridge(\tau)$ to find
\begin{equation}
\label{e:headroom_benefit}
\begin{split}
\mathrm{Additional~Headroom} = & \left (  \max_{\phi,\,\tau} |\IloadingPhiT^{\mathrm{No\,VSC}}| \right ) \\
&- \left (\max_{\phi,\,\tau} |\IloadingPhiT^{\mathrm{With\,VSC}}| \right )\,.
\end{split}
\end{equation}

\section{Results}

To study how dc link ripple affects the benefits provided by a VSC, we use two weeks of data for a 500~kVA secondary substation based in the UK \cite{nged2026demand} (id: no. 110568). Fig.~\ref{f:input_data} shows the first full day of demand $\Idemand$, showing that the additional headroom at peak demand on this day could be 60~A (greater than 20\% headroom released). Phase angles of the demand currents have an average of 120$^{\circ}$ lag between phases, with a small random angle offset at each time $\tau$ (normally distributed with zero mean and standard deviation $\pi/15$) to account for time-varying demand power factors. The optimization problems are solved with cvxpy \cite{agrawal2018rewriting, diamond2016cvxpy}.

\begin{figure}
\centering
\includegraphics[width=0.48\textwidth]{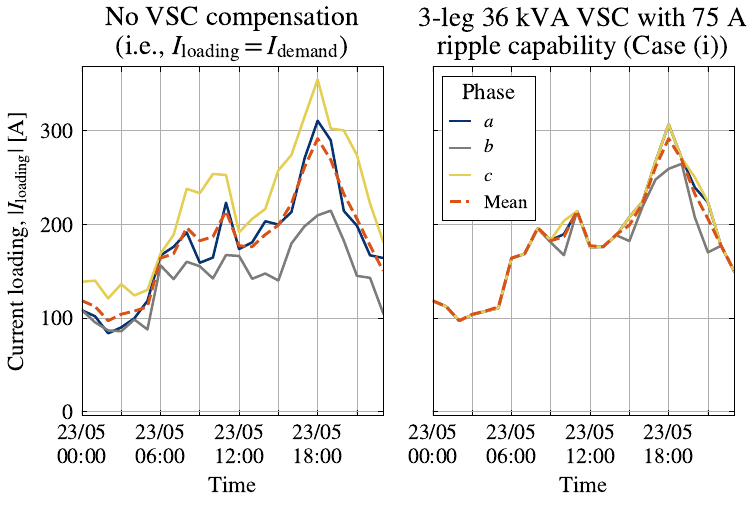}
\caption{Loading $\Iloading$ for the first full day of data, in the case of no VSC injections (left), and a 36~kVA 3-leg VSC with split dc link (Case (i)) and a 75~Arms capacitor ripple current capabilities (right).}\label{f:input_data}
\end{figure}

To study the impact of dc link ripple constraints, we consider three-, four-, and relaxed reconfigurable-leg cases (Section~\ref{ss:vsc_operational_constraints}) across VSCs with total leg capacities $\IhalfBridgeTotal$ of:
\begin{itemize}
\item 150~A, with split dc link (Case (i), 36~kVA);
\item 150~A, with monolithic dc link (Case (ii), 36~kVA);
\item 30~A, with split dc link (Case (iii), 7.2~kVA).
\end{itemize}
An example of the solution of \eqref{e:objective} across time for Case (i) is shown in Fig.~\ref{f:plt3legVsc}. This shows how a case with no ripple capabilities (left subfigure) is restricted to inject only balanced currents, as the VSC does not have the ripple capacity required to inject zero- or negative-sequence currents. In contrast, when there is a large dc link ripple capability (right subfigure), the VSC is able to inject unbalanced currents that track the unbalance seen in the load. As a result, when the unbalanced load peaks at 18:00, the VSC can draw power from phase $b$ and inject it into phase $c$, effectively mitigating the high currents, as in Fig.~\ref{f:input_data} (right subfigure).

\begin{figure}
\centering
\includegraphics[width=0.481\textwidth]{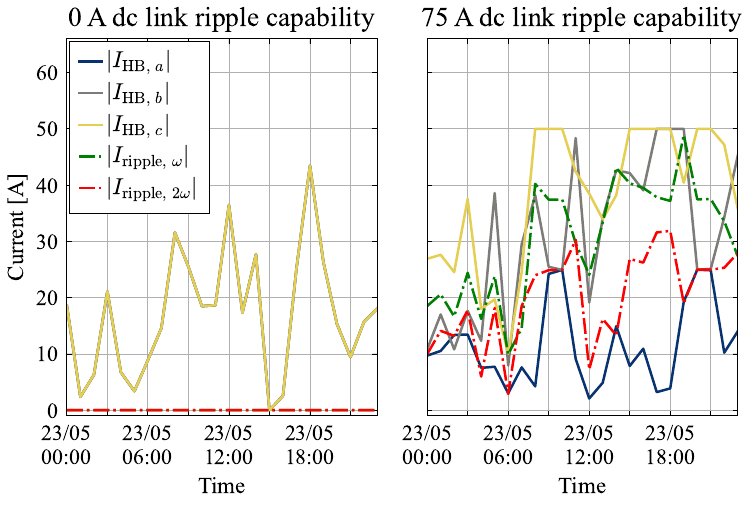}
\caption{Injected currents for a 36~kVA three-leg VSC with split dc link (Case (i)), considering 0~A dc link ripple current capabilities (left) vs full dc link ripple capability of 75~A (right).}\label{f:plt3legVsc}
\end{figure}

\subsection{Benefits for a Large VSC with Split DC Link (Case (i))}

Fig.~\ref{f:pltRippleSweepSplitPhase80} shows the Benefit \eqref{e:benefit} and Additional Headroom \eqref{e:headroom_benefit} for the full range of dc link ripple capabilities \eqref{e:ripple_sweep_parameters}. It can be seen that, as expected, the benefits increase monotonically as the dc link ripple capability $\iRippleMax$ increases. Furthermore, the change in capability plateaus as this value reaches the maximum ripple value of 75~A. Across all VSC configurations of Case (i), the dc link ripple $\iRippleMax$ more than doubles the Additional Headroom benefit, showing how small dc link ripple capabilities severely restrict the benefits of the VSC.

\begin{figure}
\centering
\includegraphics[width=0.481\textwidth]{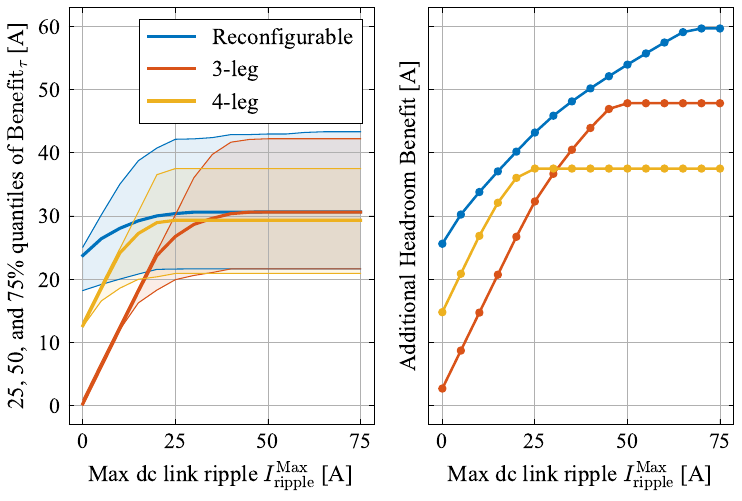}
\caption{Results for parameter sweep of the maximum dc link ripple $\iRippleMax $ for Case (i), considering quartiles of the Benefit \eqref{e:benefit} across time $\tau$ (left) and for the Additional Headroom \eqref{e:headroom_benefit} across all $\tau $ (right).}\label{f:pltRippleSweepSplitPhase80}
\end{figure}

Comparing the three half-bridge configurations (Section~\ref{ss:vsc_operational_constraints}), it can be seen that, as expected, the relaxed reconfigurable VSC topology has the greatest Benefit and Additional Headroom. The 3-leg converter initially has low benefits, as it can only inject balanced power, whilst the 4-leg converter can use the neutral-connected leg to inject zero sequence current. However, as the dc link ripple capability $\iRippleMax$ increases, the benefits of the 3-leg VSC surpass the 4-leg design, as the 3-leg VSC can exploit the dc link midpoint for neutral return (noting the 50~A legs for 3-leg design, versus 37.5~A for the 4-leg design).

\subsection{Benefits for a Large VSC with Monolithic DC Link (Case (ii)) and Small VSC with Split DC Link (Case (iii))}

Qualitatively, results for Case (ii) mirror those of Case (i), as shown in Fig.~\ref{f:pltRippleSweepSingleCap80}. The main difference between is that the Additional Headroom is somewhat reduced, and that the dc link capacitor ripple $\iRippleMax $ does not increase beyond 50~A. This is because there is no neutral return for a monolithic dc link capacitor, so the applicable upper capacitor limit is limited by \eqref{e:monolithic_cap_ripple} (i.e., 50~A) rather than \eqref{e:ripple_sweep_parameters}. The difference between the minimum and maximum Additional Headroom is smaller than Case (i), but still 80\% or greater across the three half-bridge configurations.

\begin{figure}
\centering
\includegraphics[width=0.46\textwidth]{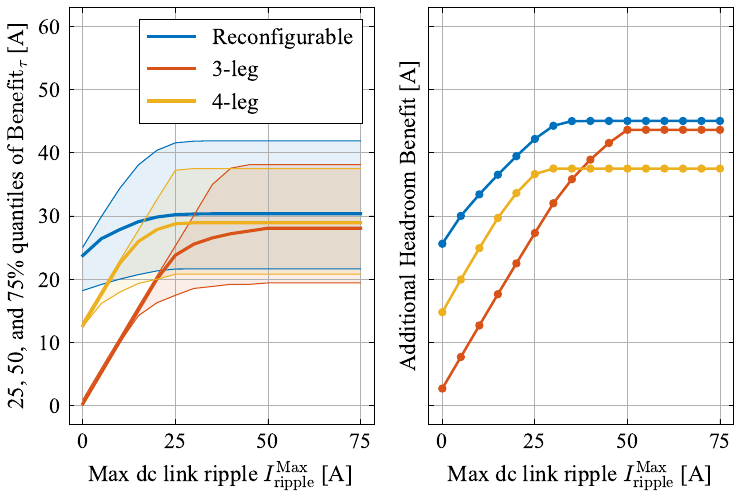}
\caption{Results for parameter sweep of the maximum dc link ripple $\iRippleMax $ for Case (ii), considering quartiles of the Benefit \eqref{e:benefit} across time $\tau$ (left) and for the Additional Headroom \eqref{e:headroom_benefit} across all $\tau $ (right).}\label{f:pltRippleSweepSingleCap80}
\end{figure}

Finally, Fig~\ref{f:pltRippleSweepSplitPhase20} shows the results for Case (iii), whereby the VSC capacity on each leg is significantly smaller than the unbalance in the network. As a result, all three half-bridge configurations show benefits that show a smaller spread across the time-varying benefits (left subfigure), as the VSC saturates its outputs at most time periods. In this case, the Additional Headroom benefit (right subfigure) shows an even greater difference in benefits than either of the other two Cases.

\begin{figure}
\centering
\includegraphics[width=0.46\textwidth]{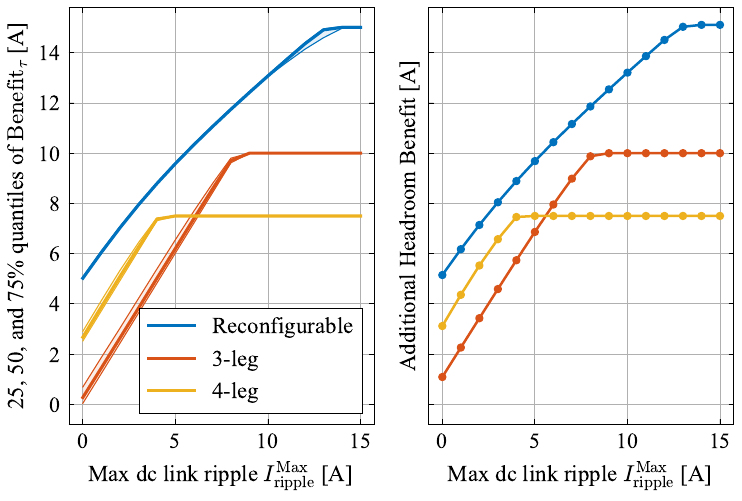}
\caption{Results for parameter sweep of the maximum dc link ripple $\iRippleMax $ for Case (iii), considering quartiles of the Benefit \eqref{e:benefit} across time $\tau$ (left) and for the Additional Headroom \eqref{e:headroom_benefit} across all $\tau $ (right).}\label{f:pltRippleSweepSplitPhase20}
\end{figure}

\section{Discussion and Conclusions}

Conventional VSC models only consider per-phase current and power constraints \eqref{e:i_leg_max}, \eqref{e:p_balance}, therefore requiring only two model parameters. In this work, we provide evidence that neglecting capacitor ripple constraints risks an inadequate model for VSCs injecting unbalanced currents, with the \emph{minimum} difference in converter overall benefits across nine topologies having value of 80\%. This suggests that these more detailed converter models are necessary.

Nevertheless, the assumptions considered in this work could be further studied to consider how they interact with the ripple constraints considered. This includes dc link voltage ripple; the effect dc link-connected active power sources; impacts of different output filter topologies; or alternative topologies such as multilevel converters. This would enable a more comprehensive study of the tradeoff between complexity and fidelity for different applications.

Conversely, utilities could develop libraries of example mission profiles for these emerging use-cases, which are non-trivial due to these converter's multiport structure. These would enable power converter topology and design to be co-optimized, improving competitiveness against alternative solutions. We conclude that utility-owned power converters are of particular interest (as compared to customer-owned converters), given their topology, design, and control can be influenced by system operators, ultimately enabling detailed converter models to be accurate and of high value.

\section*{AI Usage Disclosure Statement}
AI was not used in the research or writing of this paper.

\bibliographystyle{ieeetr}
\bibliography{master_bib_230918_ripple_opf_251229}

\endgroup
\end{document}